\UseRawInputEncoding
\documentclass[aps,onecolumn,10pt]{revtex4}
\usepackage{amsmath}
\usepackage{amssymb}
\usepackage{hyperref}
\hypersetup{colorlinks=true} 
\numberwithin{equation}{section}
\numberwithin{equation}{section}

\usepackage{graphicx}
\usepackage{epsfig}
\usepackage{float}
\begin{document}
\allowdisplaybreaks
\setcounter{equation}{0}

\title{Antilinear Symmetry and the Ghost Problem in Quantum Field Theory}

\author{Philip D. Mannheim}
\affiliation{Department of Physics, University of Connecticut, Storrs, CT 06269, USA \\
 philip.mannheim@uconn.edu\\ }

\date{April 16 2021}

\begin{abstract}

The recognition that the eigenvalues of a non-Hermitian Hamiltonian could all be real if the Hamiltonian had an antilinear symmetry such as $PT$ stimulated new insight into the underlying structure of quantum mechanics. Specifically, it led to the realization that Hilbert space could be richer than the established Dirac approach of constructing inner products out of ket vectors and their Hermitian conjugate bra vectors. With antilinear symmetry one must instead build inner products out of ket vectors and their antilinear conjugates, and it is these inner products that would be time independent in the non-Hermitian but antilinearly symmetric case even as the standard Dirac inner products would not be. Moreover, and in a sense quite remarkably, antilinear symmetry could address  not only the temporal behavior of the inner product but also  the issue of its overall sign, with antilinear symmetry being capable of yielding a positive inner product in situations such as fourth-order derivative quantum field theories where the standard Dirac inner product is found to have ghostlike negative signature. Antilinear symmetry thus solves the ghost problem in such theories by showing that they are being formulated in the wrong Hilbert space, with antilinear symmetry providing a Hilbert space that is ghost free. Antilinear symmetry does not actually get rid of the ghost states. Rather, it shows that the reasoning that led one to think that ghosts were present in the first place is faulty.

\end{abstract}

\maketitle

\section{Introduction}
\label{S1}

Interest in non-Hermitian Hamiltonians was sparked by the work of Bender and collaborators \cite{Bender1998,Bender1999} on the non-Hermitian Hamiltonian $H=p^2+ix^3$, as it turned out to have a completely real energy eigenspectrum. Ever since the work of Dirac the requirement of Hermiticity of the Hamiltonian had been foundational for quantum mechanics. However quantum mechanics did not actually need to rely on Hermiticity because it is only a sufficient condition for the reality of the eigenvalues of a Hamiltonian  but not a necessary one. Thus while Hermiticity implies reality there is no converse theorem that says that if a Hamiltonian is not Hermitian then its eigenvalues cannot all be real. However, by and large the identification of a necessary condition had not been pursued prior to the work of \cite{Bender1998,Bender1999}, and there was even a reason for this, since within linear algebra no such condition can actually be found. This is not surprising since when the necessary condition was eventually identified it turned out not to be a linear condition at all. Rather, the necessary condition is that the Hamiltonian must possess an antilinear symmetry, viz. a symmetry that does not just act on operators but one which conjugates complex numbers as well. 
 
To see how antilinearity works it is instructive to look at the eigenvector equation 
\begin{eqnarray}
i\frac{\partial}{\partial t}|\psi(t)\rangle=H|\psi(t)\rangle=E|\psi(t)\rangle.
\label{1.1g}
\end{eqnarray}
On replacing the parameter $t$ by $-t$ and then multiplying by a general antilinear operator $A$ we obtain
\begin{eqnarray}
i\frac{\partial}{\partial t}A|\psi(-t)\rangle=AHA^{-1}A|\psi(-t)\rangle=E^*A|\psi(-t)\rangle.
\label{1.2g}
\end{eqnarray}
From (\ref{1.2g}) we see that if $H$ has an antilinear symmetry so that $AHA^{-1}=H$, then, as first noted by Wigner in his study of time reversal invariance, energies can either be real and have eigenfunctions that obey $A|\psi(-t)\rangle=|\psi(t)\rangle$, or can appear in complex conjugate pairs that have conjugate eigenfunctions ($|\psi(t)\rangle \sim \exp(-iEt)$ and $A|\psi(-t)\rangle\sim \exp(-iE^*t)$). To establish a converse, suppose we are given that the energy eigenvalues are real or appear in complex  conjugate pairs. In such a case not only would $E$ be an eigenvalue but $E^*$ would be too. Hence, we can set $HA|\psi(-t)\rangle=E^*A|\psi(-t)\rangle$ in (\ref{1.2g}), and obtain
\begin{eqnarray}
(AHA^{-1}-H)A|\psi(-t)\rangle=0.
\label{1.3g}
\end{eqnarray}
Then if the eigenstates of $H$ are complete, (\ref{1.3g}) must hold for every eigenstate, to yield $AHA^{-1}=H$ as an operator identity, with $H$ thus having an antilinear symmetry. 

The above analysis thus demonstrates that antilinearity is necessary  to ensure that all energies are real or that some or all of them appear in complex conjugate pairs, and without it it is guaranteed that not all of the eigenvalues could be real. Once given antilinearity, to enforce reality alone we note that if $|\psi(-t)\rangle$ is also an eigenstate of $A$ so that up to an overall phase $A|\psi(-t)\rangle=|\psi(t)\rangle$, then from (\ref{1.1g}) and (\ref{1.2g}) it follows that $E=E^*$, while if $A|\psi(-t)\rangle\neq |\psi(t)\rangle$ it follows that $E\neq E^*$.  As discussed in \cite{Mannheim2018a}, the necessary and sufficient condition for the reality of the energy eigenvalues then is that $H$ have an antilinear symmetry and that all of  its eigenstates be eigenstates of the antilinear operator. 

Returning now to $H=p^2+ix^3$ itself, we note that it is left invariant under a particular antilinear symmetry, namely $PT$ symmetry, where $P$ is the parity operator (a linear operator) while $T$ is the antilinear time-reversal operator. (Under $PT$ $p\rightarrow p$, $x\rightarrow -x$ and $i\rightarrow -i$.) It was the realization \cite{Bender1998, Bender1999} that $PT$ symmetry underlay the $H=p^2+ix^3$ theory that opened up the study of $PT$ symmetric systems, with the field having then grown rapidly (see e.g. \cite{Bender2007,Special2012,Theme2013,ElGanainy2018,Bender2018,Fring2021} and references therein.)

Once one considers non-Hermiticity one is no longer able to use the Dirac inner product, viz. the overlap of a ket vector with a bra vector that is constructed as the Hermitian conjugate of the ket vector, as the Dirac inner product is no longer time independent (viz. $\langle \psi(t)|\psi(t)\rangle =\langle \psi(0)|e^{iH^{\dagger}t}e^{-iHt}|\psi(0)\rangle$ is not equal to $\langle \psi(0)|\psi(0)\rangle$  if $H^{\dagger}\neq H$).  Rather, one should use the overlap of a ket vector with a bra vector that is constructed as the conjugate of the ket vector with respect to the antilinear symmetry, as this then does give a time independent inner product if $H$ has an antilinear symmetry. As discussed in \cite{Mostafazadeh2002,Solombrino2002,Mannheim2018a,Mannheim2018c} this particular inner product is also equal to the overlap $\langle \psi(t)|V|\psi(t)\rangle $ where the $V$ operator effects the pseudo-Hermitian condition 
\begin{eqnarray} 
VHV^{-1}=H^{\dagger}.
\label{1.4g}
\end{eqnarray}
That this condition is equivalent to antilinear symmetry  can be seen from the fact that under it $H$ and $H^{\dagger}$ possess a common set of eigenvalues, and thus all the eigenvalues of $H$ are either real or in complex conjugate pairs, i.e., precisely the same outcome as when $H$ has an antilinear symmetry. With $|\psi(t)\rangle $ and $\langle \psi(t)|$ obeying $i\partial_t|\psi(t)\rangle=H|\psi(t)\rangle$, $-i\partial_t\langle \psi(t)|=\langle \psi(t)|H^{\dagger}$ one can immediately show that 
\begin{eqnarray}
i\frac{\partial}{\partial t} \langle \psi(t)|V|\psi(t)\rangle
=\langle \psi(t)|(VH-H^{\dagger}V)|\psi(t)\rangle.
\label{1.5g}
\end{eqnarray}
Thus as long as $V$ is invertible the $\langle \psi(t)|V|\psi(t)\rangle$ inner product is time independent when (\ref{1.4g}) is imposed, just as required.

With the recognition that with antilinearity a Hamiltonian can have real eigenvalues even if it is not Hermitian, and that with antilinearity one can have a time-independent inner product even if the Hamiltonian is not Hermitian, we turn now to the third key aspect of antilinearity, namely the role of boundary conditions. Once  we have done this we can address the issue of the ghost problem associated with the sign of the inner product.

\section{What is meant by Hermiticity}
\label{S2}

To understand the role of boundary conditions, we note that for a second-order differential operator $D$ in the self-adjoint form $D=-p(x)d^2/dx^2-p^{\prime}(x)d/dx+q(x)$ that acts on wave functions $\phi(x)$, $\psi(x)$, one can show (Green's theorem) that
\begin{eqnarray}
\int_a^bdx [\phi^*D\psi-(\psi^*D\phi)^*]=\int_a^bdx [\phi^*D\psi-[D\phi]^*\psi]=\bigg{[}p\psi\phi^{*\prime}-p\phi^*\psi^{\prime}\bigg{]}\bigg{|}_a^b.
\label{2.1g}
\end{eqnarray}
Hermiticity in the form $\langle \phi|D|\psi\rangle=\langle \psi|D|\phi\rangle^*$
is then only secured by the vanishing of the surface term at the spatial boundary. While quantum mechanics is based on commutators of operators, they are not defined until a basis on which the operators act is specified. For such a basis to be viable it needs to be well-behaved at the spatial boundary
Thus for the $[\hat{x},\hat{p}]=i$ commutator, we can only set $\hat{p}=-id/dx$ when it acts on a good, i.e., normalizable, test function according to
\begin{eqnarray}
\left[ \hat{x},-i\frac{d}{dx}\right]\psi(x)=i\psi(x).
\label{2.2g}
\end{eqnarray}
Thus for a harmonic oscillator $\hat{H}=\hat{p}^2+\hat{x}^2$ for instance we have the following two typical solutions:
\begin{eqnarray}
\left[-\frac{d^2}{dx^2}+x^2\right]e^{-x^2/2}= e^{-x^2/2},~~~\left[-\frac{d^2}{dx^2}+x^2\right]e^{+ x^2/2}=-e^{+ x^2/2},
\label{2.3gf}
\end{eqnarray}
with the eigenvalue of $e^{-x^2/2}$ being positive and the eigenvalue of $e^{+x^2/2}$ being negative. Of these two solutions
only the $e^{-x^2/2}$ wave function is normalizable (cf. vanishing of the surface term), with $\int dx \psi^*(x)\psi(x)$ being finite. And when acting on it we can indeed represent $\hat{p}$ as $\hat{p}=-id/dx$. Here $x$ is real and we are working in the coordinate basis in which $\hat{x}$ is Hermitian, has real eigenvalues $x$, and is diagonal in this basis.

But what of the $e^{+x^2/2}$ solution? It is not normalizable and we cannot represent $\hat{p}$ as $-id/dx$ when acting on it since we cannot throw away the surface term in an integration by parts. However, that does not mean that we can ignore this solution, it means only that when acting on it one cannot represent $\hat{p}$ as $-id/dx$. So maybe we could represent $\hat{p}$ by some alternate form. Thus suppose we make $x$ pure imaginary. Then $e^{+x^2/2}$
is normalizable on the imaginary axis. Thus we can take both $\hat{x}$ and $\hat{p}$ to be anti-Hermitian and represent $[\hat{x},\hat{p}]=i$ as $[-i\hat{x},i\hat{p}]=i$. This is equivalent to implementing the similarity transformation $\hat{S}=\exp(\pi \hat{p}\hat{x}/2)$ that effects
\begin{eqnarray}
\hat{S}\hat{x}\hat{S}^{-1}=-i\hat{x}=\hat{y},\quad \hat{S}\hat{p}\hat{S}^{-1}=i\hat{p}=\hat{q},
\label{2.4g}
\end{eqnarray} 
with such a similarity transformation preserving both the commutation relation $[\hat{x},\hat{p}]=[\hat{y},\hat{q}]=i$ and the eigenvalues of any Hamiltonian $\hat{H}(\hat{x},\hat{p})$ that is built out of $\hat{x}$ and $\hat{p}$. We thus have 
\begin{eqnarray}
\left[\hat{y},\hat{q}\right]\psi(y)=\left[\hat{y},-i\frac{d}{dy}\right]\psi(y)=i\psi(y),
\label{2.5g}
\end{eqnarray}
and now $e^{+x^2/2}=e^{-y^2/2}$ is a good test function. Thus $e^{-x^2/2}$ is a good test function when $x$ is real, while $e^{+x^2/2}$ is a good test function when $x$ is pure imaginary. When $x$ is pure imaginary we can set 
\begin{eqnarray}
[\hat{p}^2+\hat{x}^2]e^{+x^2/2}=-[\hat{q}^2+\hat{y}^2]e^{-y^2/2}=\left[\frac{d^2}{dy^2}-y^2\right]e^{-y^2/2}=-e^{-y^2/2}.
\label{2.6g}
\end{eqnarray}
Thus while the eigenvalues of $\hat{p}^2+\hat{x}^2$ would be positive if $\hat{p}$ and $\hat{x}$ are both Hermitian, the eigenvalues of $\hat{p}^2+\hat{x}^2$ would be negative if $\hat{p}$ and $\hat{x}$ are both anti-Hermitian.

Now taking $\hat{x}$ and $\hat{p}$ to be anti-Hermitian might not be conventional, but to take these two operators to be Hermitian in the first place has a hidden assumption, namely that they are acting on their own eigenstates. However, we are not interested in how they act on their own eigenstates, since  in order to determine the eigenstates and eigenvalues of a Hamiltonian $\hat{H}$ that is built out of them what matters  is how $\hat{x}$ and $\hat{p}$ act on the eigenstates of $\hat{H}$. Thus while both $\hat{x}$ and $\hat{p}$ might be Hermitian when acting on their own eigenstates that does not make them Hermitian when acting on the eigenstates of $\hat{H}$. This then is the secret of $PT$ symmetry, with there potentially being a mismatch between the eigenstates of a Hamiltonian and the eigenstates of the operators out of which it is built. 

To be general we note that quantum mechanics allows us to make a similarity transformation through any angle of the form  $\hat{S}=\exp(-\theta \hat{p}\hat{x})$, to thus  effect 
\begin{eqnarray}
\hat{S}\hat{x}\hat{S}^{-1}=\hat{x}\exp(i\theta) ,\quad \hat{S}\hat{p}\hat{S}^{-1}=\hat{p}\exp(-i\theta)
\label{2.7g}
\end{eqnarray}
while preserving the $[\hat{x},\hat{p}]$ commutator.
Ordinarily this is not of any significance since we work with Hermitian operators that have normalizable wave functions on the real axis, and we have no need to go into the complex plane. But if the wave functions are not normalizable on the real axis, we may be able to continue into a so-called ``Stokes wedge" in the complex plane where they then are normalizable, and cross over a ``Stokes line" that divides the two regions ($\theta=\pi/4$ in the harmonic oscillator case). The two sides of the Stokes line represent different, inequivalent, realizations of the quantum theory. And not only that, they represent formulations in different Hilbert spaces. Specifically, for the harmonic oscillator with Hamiltonian and commutator
\begin{eqnarray}
\hat{H}=a^{\dagger}a+\frac{1}{2},\quad [a,a^{\dagger}]=1,
\label{2.8g}
\end{eqnarray}
the positive energy sector is associated with a Hilbert space that is built on a vacuum that obeys $a|\Omega_1\rangle=0$, while the negative energy sector is associated with a Hilbert space that is built on a vacuum that obeys $a^{\dagger}|\Omega_2\rangle=0$.

For the $PT$ symmetry prototype $\hat{H}=\hat{p}^2+i\hat{x}^3$, we note that its energy eigenvalues would indeed be complex  if the eigenstates of $\hat{x}$ were in the same Hilbert space as the eigenstates of $\hat{H}$. However, it turns out \cite{Bender1998,Bender1999,Bender2007,Bender2018} that the eigenstates of $\hat{H}$ are not normalizable on the real $x$ axis. Nonetheless,  one can find appropriate Stokes wedges in the complex $x$ plane in which these eigenfunctions are normalizable. And in these wedges the eigenvalues of $\hat{H}$ are all real. It is in this way then that $PT$ symmetry makes $\hat{H}=\hat{p}^2+i\hat{x}^3$ viable. At the same time it serves a warning: one cannot determine an energy eigenspectrum by inspection. One has to first identify appropriate boundary conditions.

In fact $PT$ symmetry serves a second warning as well. One cannot even tell if a Hamiltonian is Hermitian by inspection. Specifically, consider as an example the relativistic action
\begin{eqnarray}
I_{\rm S}=\displaystyle{\int}d^4x \left[\frac{1}{2}\partial_{\mu}\phi\partial^{\mu}\phi-\frac{1}{2}m^2\phi^2\right]
\label{2.9g}
\end{eqnarray}
with Hamiltonian
\begin{eqnarray}
H=\displaystyle{\int } d^3x \frac{1}{2}[\dot{\phi}^2+\vec{\nabla}\phi\cdot \vec{\nabla}\phi+m^2\phi^2].
\label{2.10g}
\end{eqnarray}
Solutions to the  wave equation  obey 
\begin{eqnarray}
\phi(\vec{x},t)=\displaystyle{\sum} [a(\vec{k})\exp(-i\omega_k t+i\vec{k}\cdot\vec{x})+a^{\dagger}(\vec{k})\exp(+i\omega_k  t-i\vec{k}\cdot\vec{x})],~~~~~\omega^2_k=\vec{k}^2+m^2,
\label{2.11g}
\end{eqnarray}
 and the Hamiltonian is given by 
 \begin{eqnarray}
 H=\displaystyle{\sum}\frac{1}{2}[\vec{k}^2+m^2]^{1/2} [a^{\dagger}(\vec{k})a(\vec{k})+a(\vec{k})a^{\dagger}(\vec{k})].
 \label{2.12g}
 \end{eqnarray}
 If $m^2>0$ all energies are real, and both $H$ and $\phi(\vec{x},t)$ are Hermitian. However, if $m^2=-n^2<0$, then 
 \begin{eqnarray}
 \omega^2_k=\vec{k}^2-n^2.
 \label{2.13g}
 \end{eqnarray}
Thus now the $k<n$ energies come in complex conjugate pairs, and neither $H$ nor $\phi(\vec{x},t)$ is Hermitian. And yet inspection of (\ref{2.9g}) would have suggested that even for negative $m^2$ both $H$ and $\phi(\vec{x},t)$ are Hermitian.
 
As a second example consider the non-relativistic Hamiltonian  
\begin{eqnarray}
H_{\rm PU}(\alpha,\beta)=\frac{p_x^2}{2}+p_zx+(\alpha^2-\beta^2)x^2-\frac{1}{2}(\alpha^2+\beta^2)^2z^2
\label{2.14g}
\end{eqnarray}
with real $\alpha$ and $\beta$.
This Hamiltonian is associated with the Pais-Uhlenbeck ($PU$) fourth-order derivative oscillator theory \cite{Pais1950} that we discuss below. This Hamiltonian was constructed in \cite{Mannheim2005} and explored in detail in \cite{Bender2008a,Bender2008b}.  
It also appears to be Hermitian. However, as we show below,  the energy eigenvalues of its two first excited states are the complex $\alpha+i\beta$ and $\alpha-i\beta$. Thus to determine if a Hamiltonian is indeed Hermitian one needs to solve the theory and get the eigenvalues first.  

As a third example consider the non-relativistic Hamiltonian  
\begin{eqnarray}
H_{\rm PU}(\omega)=\frac{p_x^2}{2}+p_zx+\omega^2x^2-\frac{1}{2}\omega^4z^2,
\label{2.15g}
\end{eqnarray}
with real $\omega$.
It also appears to be Hermitian. However, as we show below,  the Hamiltonian only has a single one-particle state, with energy equal to the  $2\omega$. And while this energy eigenvalue is even real, the Hamiltonian actually turns out to a non-diagonalizable Jordan-block Hamiltonian as there is only one first excited state, with there not being any second eigenvector, to thus leave the Hamiltonian with an incomplete set of eigenvectors. With the Hamiltonian not being diagonalizable it cannot be Hermitian, and cannot be even though its eigenvalues are real. Thus to determine if a Hamiltonian is indeed Hermitian one needs to solve the theory and show that the set of eigenvectors is complete. 

As a fourth example consider the non-relativistic Hamiltonian
\begin{eqnarray}
H_{\rm PU}(\omega_1,\omega_2)=\frac{p_x^2}{2}+p_zx+\frac{1}{2}\left(\omega_1^2+\omega_2^2 \right)x^2-\frac{1}{2}\omega_1^2\omega_2^2z^2
\label{2.16g}
\end{eqnarray}
with real $\omega_1>\omega_2>0$.
Now for this example, while the energy eigenvalues of its two first excited states are the real $\omega_1$ and $\omega_2$, and while this time the set of energy eigenvectors is complete, its eigenfunctions are not normalizable on the real axis. And so yet again it is not Hermitian. However, as we discuss below, there does exist a Stokes wedge in which its eigenvectors are normalizable \cite{Bender2008a}, and this wedge does not contain the real axis.

While it is not Hermiticity that these four examples have in common, there is some other attribute that they do share.  In the first two of the cases one has a complex conjugate pair of energy eigenvalues. Thus both Hamiltonians must have an antilinear symmetry. In \cite{Bender2008a,Bender2008b} it was shown that the third and fourth examples have an antilinear symmetry too. Thus all four of the  Hamiltonians have an antilinear symmetry, with it thus being antilinearity rather than Hermiticity that they all have in common. Now in \cite{Mannheim2018a} it was shown on general grounds that under only two requirements, namely invariance under the complex Lorentz group and conservation of probability, the Hamiltonian must be CPT invariant where $C$ is the charge conjugation operator. And it must be so regardless of whether or not it might be Hermitian. (The standard proof of the $CPT$ theorem required Hermiticity.)  Now while one cannot  a priori tell whether a Hamiltonian might be Hermitian, one can tell a priori if it has an antilinear symmetry as that requires no need to solve the theory, and is fixed solely by the behavior of the operators and coefficients that appear in the Hamiltonian. Now for our four examples $C$ just happens to be separately conserved, and thus for all of them $CPT$ reduces to $PT$. In fact this is a general requirement for any non-relativistic theory that descends from a covariant relativistic theory (which any  theory must do if it is  to be physically relevant). Thus,  if one is below the threshold for pair creation $C$ is separately conserved and $CPT$ defaults to $PT$. This thus puts $PT$ symmetry on a quite secure theoretical footing.

Both (\ref{2.14g}) and (\ref{2.15g}) can be derived by setting $\omega_1=\alpha+i\beta$, $\omega_2=\alpha-i\beta$ and $\omega_1=\omega_2=\omega$ in (\ref{2.16g}). Now in Sec. \ref{S1} we had noted that with antilinearity one could have real eigenvalues or eigenvalues that are in a complex conjugate pair. However, we also noted that in the complex pair realization the eigenvectors are complex conjugates of each other. Thus if we switch off the complex part by setting $\beta$ to zero in (\ref{2.14g}) the eigenvalues not only become equal, so do the eigenvectors. Thus in this limit we must lose an eigenvector. Thus the transition from (\ref{2.16g}) to (\ref{2.14g}) must go through a singular point (known in the $PT$ literature as an exceptional point) in which the Hamiltonian becomes of non-diagonalizable Jordan-block form. Thus for antilinear symmetry there are actually three realizations: eigenvalues all real and eigenvectors complete, eigenvalues all real and but eigenvectors incomplete, eigenvalues in complex conjugate pairs and eigenvectors complete. As discussed in \cite{Bender2008a,Bender2008b,Mannheim2018a}, in all cases one can construct a time-independent probability.

Turning back now to the harmonic oscillator, we note that while we can find a Stokes wedge in which its $e^{+x^2/2}$ type eigenfunctions are normalizable, no amount of similarity transforming can convert  the associated negative energy eigenvalues into positive ones. And it is for this reason that we have to abandon this sector of the oscillator theory. For our purposes here we note that there is a second disease in this sector, namely with $a^{\dagger}$ annihilating the vacuum according to $a^{\dagger}|\Omega_2\rangle=0$, it is the operator $a$ that serves as the creator, with the one-particle matrix element $\langle \Omega_2|a^{\dagger}a|\Omega_2\rangle$ then being of ghostlike negative signature. As we now show, a similar ghost problem occurs for the $PU$ oscillator as well, but using the techniques of $PT$ symmetry that we have described above, this ghost problem proves to be solvable.

\section{The Ghost Problem of Fourth-Order Derivative Theories}
\label{S3}

In quantum field theory radiative corrections generate ultraviolet divergences. In renormalizable field theories such as quantum electrodynamics one needs to regularize these infinities, while in non-renormalizable theories such as Einstein gravity one tries to augment the theory so as to lower the degree of divergence.  The prototype for regularizing divergences in quantum field theories while maintaining  gauge invariance is the Pauli-Villars regularization scheme proposed in \cite{Pauli1949} in which one replaces the standard one-particle $1/(k^2-M^2)$ propagator by the two-particle  
\begin{eqnarray}
D_{\rm{PV}}(k)=\frac{1}{k^2-M_1^2}-\frac{1}{k^2-M_2^2},
\label{3.1g}
\end{eqnarray} 
so that a $1/k^2$ asymptotic ultraviolet behavior at large $k^2$ would be replaced by the more convergent $1/k^4$. 
As conceived in \cite{Pauli1949}, the two particles would be associated with a second-order derivative action of the form
\begin{eqnarray}
I_{S_1}+I_{S_2}=\int d^4x\left[\tfrac{1}{2}\partial_{\mu}\phi_1\partial^{\mu}\phi_1-\tfrac{1}{2}M_1^2\phi_1^2\right]+\int d^4x\left[\tfrac{1}{2}\partial_{\mu}\phi_2\partial^{\mu}\phi_2-\tfrac{1}{2}M_2^2\phi_2^2\right],
\label{3.2g}
\end{eqnarray}
with the associated wave operators requiring the $D_{\rm PV}$ propagator to obey
\begin{eqnarray}
(\partial_t^2-\nabla^2+M_1^2)\langle\Omega_1|T[\phi_1(x)\phi_1(0)]|\Omega_1\rangle=-\delta^4(x),\quad (\partial_t^2-\nabla^2+M_2^2)\langle\Omega_2|T[\phi_2(x)\phi_2(0)]|\Omega_2\rangle=+\delta^4(x).
\label{3.3h}
\end{eqnarray}
If we apply the wave operator $(\partial_t^2-\nabla^2+M_2^2)$ to $\langle\Omega_2|T[\phi_2(x)\phi_2(0)]|\Omega_2\rangle$ we generate a term of the form $\langle\Omega_2|\delta(t)[\dot{\phi}_2(x),\phi_2(0)]|\Omega_2\rangle$. With the conjugate of $\phi_2$ being $\delta I_{S_2}/\delta\phi_2=+\dot{\phi}_2$, to recover (\ref{3.3h}) we would need to set $[\dot{\phi}_2(x,t=0),\phi_2(0)]=+\delta^3(x)$ rather than the standard  $[\dot{\phi}_1(x,t=0),\phi_1(0)]=-\delta^3(x)$ used for $\phi_1$.

To confirm this choice of quantization, we note that following the insertion of 
\begin{eqnarray}
\sum |n_1\rangle\langle n_1|-\sum |n_2\rangle\langle n_2|=I
\label{3.4h}
\end{eqnarray}
with its ghostlike relative minus sign into
\begin{eqnarray}
D_{\rm PV}(k)=\int d^4x e^{ik\cdot x}\left[\langle\Omega_1|T[\phi_1(x)\phi_1(0)]|\Omega_1\rangle+\langle\Omega_2|T[\phi_2(x)\phi_2(0)]|\Omega_2\rangle\right]
\label{3.5h}
\end{eqnarray}
we also recover (\ref{3.1g}). Thus ghost states  could reduce asymptotic divergences, but at the price of loss of probability and loss of unitarity.

In order to gain further insight into the Pauli-Villars regulator scheme Pais and Uhlenbeck \cite{Pais1950} replaced the two-field action by a one-field fourth-order derivative action 
\begin{eqnarray}
I_S&=&\tfrac{1}{2}\int d^4x\bigg{[}\partial_{\mu}\partial_{\nu}\phi\partial^{\mu}
\partial^{\nu}\phi-(M_1^2+M_2^2)\partial_{\mu}\phi\partial^{\mu}\phi
+M_1^2M_2^2\phi^2\bigg{]},
\label{3.6h}
\end{eqnarray}
with a fourth-order derivative equation of motion given by
\begin{eqnarray}
&&(\partial_t^2-\vec{\nabla}^2+M_1^2)(\partial_t^2-\vec{\nabla}^2+M_2^2)
\phi(x)=0,
\label{3.7h}
\end{eqnarray}
and an associated propagator of the same form as the Pauli-Villars propagator:
\begin{eqnarray}
D_{\rm {PU}}(k)=\frac{1}{(k^2-M_1^2)(k^2-M_2^2)}=\frac{1}{(M_1^2-M_2^2)}\left(\frac{1}{k^2-M_1^2}-\frac{1}{k^2-M_2^2}\right).
\label{3.8h}
\end{eqnarray}
If we identify the $D_{\rm {PU}}(k)$ propagator as 
\begin{equation}
D_{\rm {PU}}(k)=\int d^4x e^{ik\cdot x}\langle \Omega|T[\phi(x)\phi(0)]|\Omega\rangle,
\label{3.9h}
\end{equation}
then the insertion of (\ref{3.4h}) 
into  $\langle \Omega|T[\phi(x)\phi(0)]|\Omega\rangle$ would generate $D_{\rm {PU}}(k)$. So with the relative minus sign in (\ref{3.4h}) again there appear to be ghost states and loss of unitarity. 

Although the study of \cite{Pais1950} was motivated by the Pauli-Villars regularization scheme, in trying to construct renormalizable quantum theories of gravity one also encounters fourth-order derivative theories. Specifically, up to metric indices the action given in (\ref{3.6h}) describes the coupling of the second-order derivative Einstein theory to a fourth-order derivative theory, which unlike the Einstein theory itself, is based not on the Ricci scalar alone but also on its square. To make such theories viable one thus needs to find a way to eliminate the ghost states that they appear to possess \cite{Stelle1977}, and to do so while retaining their beneficial role for renormalizability. This we shall do below. However, before doing so we should point out that the identification of the c-number propagator given in (\ref{3.8h}) with a matrix element such as that given in (\ref{3.9h}) cannot actually be made  prior to solving the underlying quantum theory. While one can go from a quantum-mechanical Hilbert space to c-number propagators one cannot infer the structure of the underlying q-number theory from a knowledge of the  c-number propagator alone. And when we now do construct the underlying Hilbert space we  will find that the propagator given in (\ref{3.8h}) should be identified with a completely different matrix element, one that involves no Hilbert space with any ghost states at all.

\section{Quantizing the  Pais-Uhlenbeck Oscillator Theory} 
\label{S4}

Since only time derivatives are relevant for quantization,  Pais and Uhlenbeck simplified the discussion by restricting to one momentum state $k$, and  with real and positive $\omega_1=(\bar{k}^2+M_1^2)^{1/2}$ and real and positive $\omega_2=(\bar{k}^2+M_2^2)^{1/2}$ the $I_S$ action given in (\ref{3.6h}) reduces to the acceleration-dependent Pais-Uhlenbeck action
\begin{eqnarray}
I_{\rm PU}=\tfrac{1}{2}\int dt\left[{\ddot z}^2-\left(\omega_1^2
+\omega_2^2\right){\dot z}^2+\omega_1^2\omega_2^2z^2\right],
\label{4.1g}
\end{eqnarray}
with equation of motion and propagator
\begin{eqnarray}
&&\frac{d^4z}{dt^4}+(\omega_1^2+\omega_2^2)\frac{d^2z}{dt^2}+\omega_1^2\omega_2^2
z=0,
\nonumber\\
&&
G_{\rm PU}(E)=\frac{1}{(E^2-\omega_1^2)(E^2-\omega_2^2)}=\frac{1}{(\omega_1^2-\omega_2^2)}\left(\frac{1}{E^2-\omega_1^2}-\frac{1}{E^2-
\omega_2^2}\right).
\label{4.2h}
\end{eqnarray}
This action is a constrained action since with only $z$, $\dot{z}$ and $\ddot{z}$, there are too many canonical variables for one oscillator but not enough for two. So on setting set $x=\dot{z}$ and using the method of Dirac Constraints  we obtain \cite{Mannheim2005} the two-oscillator Hamiltonian  
\begin{eqnarray}
H_{\rm PU}(\omega_1,\omega_2)=\tfrac{1}{2}p_x^2+p_zx+\tfrac{1}{2}\left(\omega_1^2+\omega_2^2 \right)x^2-\tfrac{1}{2}\omega_1^2\omega_2^2z^2, \quad [z,p_z]=i, \quad [x,p_x]=i,
\label{4.3g}
\end{eqnarray}
with canonical variables $z$, $p_z$, and $x$, $p_x$. And before even quantizing the theory (i.e., before implementing the  $[z,p_z]=i$, $[x,p_x]=i$ commutation relations) we immediately encounter a separate problem in addition to the ghost problem. Specifically, because of the overall minus sign in the $-\tfrac{1}{2}\omega_1^2\omega_2^2z^2$  term, the energy spectrum  is unbounded from below. The existence of this kind of classical instability  was first identified by  Ostrogradski \cite{Ostrogradski1850} as being characteristic of higher-derivative theories. 

To now quantize the theory we make the standard substitutions
\begin{eqnarray}
z&=&a_1+a_1^{\dagger}+a_2+a_2^{\dagger},\quad p_z=i\omega_1\omega_2^2
(a_1-a_1^{\dagger})+i\omega_1^2\omega_2(a_2-a_2^{\dagger}),
\nonumber\\
x&=&-i\omega_1(a_1-a_1^{\dagger})-i\omega_2(a_2-a_2^{\dagger}),\quad
p_x=-\omega_1^2 (a_1+a_1^{\dagger})-\omega_2^2(a_2+a_2^{\dagger}),
\label{4.4g}
\end{eqnarray}
and obtain a Hamiltonian and commutator algebra \cite{Mannheim2005} 
\begin{align}
H_{\rm PU}(\omega_1,\omega_2)&=2(\omega_1^2-\omega_2^2)(\omega_1^2 a_1^{\dagger}
a_1-\omega_2^2a_2^{\dagger} a_2)+\tfrac{1}{2}(\omega_1+\omega_2),
\nonumber\\
[a_1,a_1^{\dagger}]&=\frac{1}{2\omega_1(\omega_1^2-\omega_2^2)},\quad [a_2,a_2^{\dagger}]=-\frac{1}{2\omega_2(\omega_1^2-\omega_2^2)}.
\label{4.5g}
\end{align}
We note that on taking  $\omega_1>\omega_2>0$ for definitiveness, the sign of the $[a_2,a_2^{\dagger}]$ commutator is negative.  

If we define the vacuum according to 
\begin{eqnarray}
a_1|\Omega_1\rangle=0, \quad a_2|\Omega_1\rangle=0,
\label{4.6g}
\end{eqnarray}
we find that the energy spectrum is bounded from below with $|\Omega_1\rangle$ being the ground state with energy $\left(\omega_1+\omega_2\right)/2$. But  the excited state
$a_2^\dag|\Omega_1\rangle$, which lies at energy $\omega_2$ above the ground state, has a Dirac norm $\langle\Omega_1|a_2a_2^\dag|\Omega_1\rangle$ that is negative. 

On the other hand if we define the vacuum according to 
\begin{eqnarray}
a_1|\Omega_2\rangle=0,~~~a_2^{\dagger}|\Omega_2\rangle=0,
\label{4.7g}
\end{eqnarray}
the theory is then free of negative-norm states, but the energy spectrum is unbounded from below, with negative energy states propagating forward in time. 

As we see, the discussion parallels the discussion we gave in Sec. \ref{S2} for the single harmonic oscillator as we again have to deal with negative energies or negative norms. However, unlike that case we note that now the two problems occur in different Hilbert spaces not in the same one,  one built on a vacuum that obeys $a_2|\Omega_1\rangle=0$ and the other built on a vacuum that obeys  $a_2^{\dagger}|\Omega_2\rangle=0$. While having either negative norms or negative energies would be problematic,  in no single Hilbert space does one have to deal with both problems. There is nothing one could do with a theory whose energy spectrum is unbounded from below. Thus the only chance for the theory is if we can make sense of the $a_2|\Omega_1\rangle=0$ sector with its seeming states of negative norm. As we now show, using the techniques of $PT$ theory we actually can do that, with the solution involving the one thing we have not yet discussed for the $PU$ oscillator theory, namely boundary conditions. And that $PT$ symmetry might even be relevant, we note that all the eigenvalues of $H_{\rm PU}$ are real (standard two-oscillator spectrum, with all the poles of (\ref{3.8h}) and (\ref{4.2h}) being on the real frequency axis). As our discussion in Sec. \ref{S1} indicates, the Hamiltonian is either Hermitian or has an antilinear symmetry. Boundary conditions will then tell us which one it is. 

As to whether $PT$ might actually be able to solve this ghost problem at all, we recall that it has already had success in solving another ghost problem, namely the one associated with the Lee model \cite{Lee1954}. The Lee model was itself introduced in order to address renormalization issues as in it one could determine the relation between the bare and renormalized coupling constants in a closed form. This theory also turned out to have a negative Dirac norm ghost problem, and given its relevance to Pauli-Villars this aspect of it was much studied at the time \cite{Kallen1955,Heisenberg1957}. However, it was only after the advent of $PT$ theory that this problem was finally solved \cite{Bender2005}. In the Lee model the ghost state is only found to occur for a specific range of values of the bare coupling constant, and in this range the bare coupling constant is complex. This meant that the theory was not Hermitian in this range, and the authors of \cite{Bender2005} showed that it instead was $PT$ symmetric. One then has to use the $PT$ inner product rather than the Dirac one, and the $PT$ norm was found to be positive. The resolution of this ghost problem should be regarded as a considerable triumph for $PT$ theory, especially since it had not at all been developed for that purpose. So we now apply the same techniques to the Pais-Uhlenbeck oscillator theory.

\section{Solving the Ghost Problem}
\label{S5}

To identify the appropriate boundary conditions for the Pais-Uhlenbeck theory we set $p_z=-i\partial_z$, $p_x=-i\partial_x$ in (\ref{4.3g}) and solve the associated Schr\"odinger equation. For the state $|\Omega_1\rangle$ with energy  $(\omega_1+\omega_2)/2$ we obtain the wave function \cite{Mannheim2005}
\begin{eqnarray}
\psi_0(z,x)={\rm exp}\left[\tfrac{1}{2}(\omega_1+\omega_2)\omega_1\omega_2
z^2+i\omega_1\omega_2zx-\tfrac{1}{2}(\omega_1+\omega_2)x^2\right],
\label{5.1g}
\end{eqnarray}
with analogous wave functions being found for the excited states that are built out of $|\Omega_1\rangle$.
The wave function $\psi_0(z,x)$ is a somewhat unusual wave function as it is convergent in one of its variables ($x$) while being divergent in its other one ($z$). Since it is divergent in $z$ the wave function is not normalizable on the real $z$ axis. Hence the Dirac  norm of $|\Omega_1\rangle$, viz. $\langle \Omega_1|\Omega_1\rangle=\int dzdx \langle \Omega_1|z,x\rangle\langle z,x|\Omega_1\rangle=\int dzdx\psi^*_0(z,x)\psi_0(z,x)$, is not finite. Thus even before we ask what to do about this divergence, we note immediately that the closure relation given in (\ref{3.4h}), viz. $\sum |n_1\rangle\langle n_1|-\sum |n_2\rangle\langle n_2|=I$, could not hold as its validity assumes that all the states that appear in it are normalizable. Hence any conclusions that might be drawn from it such as the existence of ghost states are not valid. This of course does not mean that there are no ghost states, it means only that this particular line of reasoning is invalid.

To address whether or not there might still be ghost states anyway, we need to deal with the divergence of $\psi_0(z,x)$ at large $z$. Noting that it would become normalizable if $z$, but not $x$, were to be taken to be  pure imaginary,  we thus identify two complex $z$ plane Stokes wedges in which $\psi_0(z,x)$ is normalizable,  namely the upper and lower quadrants of a letter $X$ shape drawn in the complex $z$ plane. The quantum theory does not live in the left and right quadrants of this letter $X$, and while the $-\tfrac{1}{2}\omega_1^2\omega_2^2z^2$ term might give an Ostrogradski instability with real $z$, it does not do so with imaginary $z$. To implement making $z$ pure imaginary we follow (\ref{2.7g}) and implement the similarity transformation 
\cite{Bender2008a}
\begin{eqnarray}
y=e^{\pi p_zz/2}ze^{-\pi p_zz/2}=-iz,\quad q=e^{\pi p_zz/2}p_ze^{-\pi p_zz/2}=
ip_z,\quad [y,q]=i,\qquad [x,p]=i,\quad p_x=p,
\label{5.2g}
\end{eqnarray}
to obtain
\begin{eqnarray}
e^{\pi p_zz/2}H_{\rm PU}(\omega_1,\omega_2)e^{-\pi p_zz/2}=\bar{H}=\frac{p^2}{2}-iqx+\frac{1}{2}\left(\omega_1^2+\omega_2^2
\right)x^2+\frac{1}{2}\omega_1^2\omega_2^2y^2.
\label{5.3g}
\end{eqnarray}
With $\psi_0(z=iy,x)$ taking the form
\begin{eqnarray}
\psi_0(y,x)&=&{\rm exp}\left[-\tfrac{1}{2}(\omega_1+\omega_2)\omega_1\omega_2
y^2-\omega_1\omega_2yx-\tfrac{1}{2}(\omega_1+\omega_2)x^2\right],
\nonumber\\
&=&{\rm exp}\left\{-\frac{1}{2(\omega_1+\omega_2)}\left[\omega_1\omega_2\left[(\omega_1+\omega_2)y+x\right]^2
+x^2(\omega_1^2+\omega_2^2+\omega_1\omega_2)\right]\right\},
\label{5.4f}
\end{eqnarray}
and with $\omega_1$ and $\omega_2$ both positive, the wave function is now well-behaved for all $x$ and $y$. With $p$, $q$, $x$ and $y$ now acting as Hermitian operators on $\psi_0(y,x)$, through its $i$ factor the Hamiltonian $\bar{H}$ is manifestly not Hermitian, but all of its eigenvalues are still real since one cannot change eigenvalues by a similarity transformation, and they already were all real before we made the transformation. Given the similarity of its $i$ dependence to that of $H=p^2+ix^3$, we find \cite{Bender2008a} 
that $\bar{H}$ thus falls into the class of non-Hermitian theories that have a PT symmetry ($x$ has $P$, $T$ and $PT$ phases $(-,+,-)$, $p$ is $(-.-.+)$, $q$ is $(+,+,+)$, $y$ is $(+,-,-)$) as realized in the PT symmetry realization in which all energy eigenvalues are real. 

With $\bar{H}$ not being Hermitian we must distinguish between its left-eigenvectors and right-eigenvectors (as defined in (\ref{5.11e}) below), and for the ground state they take the form 
\begin{eqnarray}
\psi^R_0(y,x,t)&=&{\rm exp}\left[-\tfrac{1}{2}(\omega_1+\omega_2)\omega_1\omega_2
y^2-\omega_1\omega_2yx-\tfrac{1}{2}(\omega_1+\omega_2)x^2-\tfrac{i}{2}(\omega_1+\omega_2)t\right],
\nonumber\\
\psi^L_0(y,x,t)&=&{\rm exp}\left[-\tfrac{1}{2}(\omega_1+\omega_2)\omega_1\omega_2
y^2+\omega_1\omega_2yx-\tfrac{1}{2}(\omega_1+\omega_2)x^2+\tfrac{i}{2}(\omega_1+\omega_2)t\right],
\label{5.5e}
\end{eqnarray}
with a normalization
\begin{eqnarray}
\int dx dy\psi^L_0(y,x,t)\psi^R_0(y,x,t)=\frac{\pi}{(\omega_1+\omega_2)(\omega_1\omega_2)^{1/2}}
\label{5.6e}
\end{eqnarray}
that is finite, time-independent and real.

To remove the $-iqx$ cross term from $\hat{H}$  and decouple the oscillators we  introduce a Hermitian operator $Q$ according to 
\begin{eqnarray}
Q=\alpha (pq+\omega_1^2\omega_2^2xy)=Q^{\dagger},\qquad \alpha=\frac{1}{\omega_1\omega_2}{\rm log}\left(\frac{\omega_1+\omega_2}{\omega_1-\omega_2}\right),
\label{5.7e}
\end{eqnarray}
with an additional similarity transformation bringing $\hat{H}$ to the form \cite{Bender2008a}
\begin{eqnarray}
e^{-Q/2}\bar{H}e^{Q/2}&=&\bar{H}^{\prime}
=\frac{p^2}{2}+\frac{q^2}{2\omega_1^2}+
\frac{1}{2}\omega_1^2x^2+\frac{1}{2}\omega_1^2\omega_2^2y^2.
\label{5.8e}
\end{eqnarray}
We recognize $\bar{H}^{\prime}$ as being a fully acceptable standard, positive norm  two-dimensional oscillator system. 

In addition, we note that with its phase being $-Q/2$ rather than $-iQ/2$, the $e^{-Q/2}$ operator is not unitary. The transformation from $\bar{H}$ to $\bar{H}^{\prime}$ is thus not a unitary transformation, but is a transformation from a skew basis with eigenvectors $|n\rangle$ to an orthogonal basis with eigenvectors 
\begin{eqnarray}
|n^{\prime}\rangle=e^{-Q/2}|n\rangle,\quad \langle n^{\prime}|=\langle n|e^{-Q/2}.
\label{5.9e}
\end{eqnarray}
Then since $\langle n^{\prime}|m^{\prime}\rangle =\delta_{mn}$, the eigenstates of $\bar{H}$ obey
\begin{align}
&\langle n|e^{-Q}|m\rangle=\delta_{mn},\quad \sum_n|n\rangle\langle n|e^{-Q}=I,\quad \bar{H}=\sum _n|n\rangle E_n\langle n|e^{-Q}, \quad
\bar{H}|n\rangle=E_n|n\rangle,\quad \langle n|e^{-Q}\bar{H}=\langle n|e^{-Q}E_n.
\label{5.10e}
\end{align}
We thus recognize the inner product as being not $\langle n|m\rangle$ but $\langle n|e^{-Q}|m\rangle$, with the conjugate of $|n\rangle$ being $\langle n|e^{-Q}$, where according to (\ref{5.10e}) $\langle n|e^{-Q}$ is a left eigenstate $\langle L|$ of $\bar{H}$. 

Moreover, the $e^{-Q}$ transformation effects $e^{-Q}\bar{H}e^{Q}=\bar{H}^{\dagger}$. The inner product $\langle n|e^{-Q}|m\rangle$ thus emerges as the time-independent $V$ norm $\langle n|V|m\rangle$ that we introduced in Sec. \ref{S1}, with $e^{-Q}$ serving as $V$ in the $PU$ case, with $V$ effecting $V\bar{H}V^{-1}=\bar{H}^{\dagger}$. We can thus write the norm as $\langle L|R\rangle$ where $|R\rangle$ is a right-eigenvector of $\bar{H}$, and $\langle L|=\langle R|V$ is a left-eigenvector since
\begin{equation}
i\partial_t|R\rangle=\bar{H}|R\rangle, \quad -i\partial_t\langle L|=-i\partial_t\langle R|V=\langle R|\bar{H}^{\dagger}V=\langle R|V\bar{H}=\langle L|\bar{H}.
\label{5.11e}
\end{equation}
In addition, the state $\langle n|e^{-Q}$ is also the $PT$ conjugate of $|n\rangle$, so that the inner product is the overlap of a state with its $PT$ conjugate rather than that with its Hermitian conjugate, just as we had noted earlier. And as such this inner product is positive definite since $\langle n^{\prime}|m^{\prime}\rangle =\delta_{mn}$ is. The $PU$ oscillator theory is thus a fully viable unitary theory with no states of negative norm in its Hilbert space.

However, now that we have obtained a very conventional looking $\bar{H}^{\prime}$ we have to ask what happened to the propagator and whether we may have lost the relative minus sign in  $D_{\rm PU}(k)$ and its non-relativistic limit as given in (\ref{3.8h}) and (\ref{4.2h}).  Since all we have done is make a similarity transformation we could not change the signs in the propagator. However given (\ref{5.11e}) and the discussion given above, we have to conclude that the propagator had been misidentified in setting it equal to the standard matrix element given in (\ref{3.9h}). Instead, we must associate it with
\begin{equation}
D_{\rm {PU}}(k)=\int d^4x e^{ik\cdot x}\langle \Omega|VT[\phi(x)\phi(0)]|\Omega\rangle,
\label{5.12e}
\end{equation}
with use of the relativistic generalization of  (\ref{5.10e}) for the insertion of strictly positive norm intermediate states then, as explicitly shown in \cite{Bender2008b} (and in \cite{Mannheim2018b} for the $PU$ case itself), enabling us to recover the relative minus sign in a theory possessing no states with negative norm.

In Sec. \ref{S3} we identified two different theories, one based on the two-field action $I_{S_1}+I_{S_2}$ given in (\ref{3.2g}), and the other based on the one-field $I_S$ given in (\ref{3.6h}). Both theories lead to the propagator given in (\ref{3.8h}). And since the theory based on $I_{S_1}+I_{S_2}$ explicitly contains the ghost states that appear in the closure relation given in (\ref{3.4h}) it is generally thought that the theory based on $I_S$ must have ghost states too. However, one cannot draw such a conclusion since the theories based on  $I_{S_1}+I_{S_2}$ and $I_S$ are different theories. Thus ghosts in one of the theories does not necessitate ghosts in the other one as well. And as we have seen, if we associate the propagator given in (\ref{3.8h}) with a fourth-order derivative theory rather than a second-order one, then one not only has a theory that is unitary, in addition none of its good behavior in the ultraviolet is impaired. 

To provide some further insight into our resolution of the ghost problem, we note that when we introduced creation and annihilation operators $a_1,a_2,a_1^{\dagger},a_2^{\dagger}$ for the original variables $z,p_z,x,p_x$ in (\ref{4.4g}), we were led to a matrix element $\langle\Omega_1|a_2a_2^{\dagger}|\Omega_1\rangle$ that is negative. Since such a matrix element is in the form of the modulus squared of $a_2^{\dagger}|\Omega_1\rangle$ it should be positive, and so it was proposed to resolve this conflict by quantizing the theory with an indefinite Hilbert space metric. As formulated, this analysis presupposed that the original variables $z,p_z,x,p_x$ are all Hermitian, so that $a_2^{\dagger}$ is then the Hermitian conjugate of $a_2$. However, by a study of boundary conditions we have found that this is not in fact the case. Consequently,
$a_2^{\dagger}$ is not in fact the Hermitian conjugate of $a_2$, and there is then no contradiction in obtaining a negative value for $\langle \Omega_1|a_2a_2^{\dagger}|\Omega_1\rangle$, and thus no need to invoke a negative metric at all. On constructing a creation and annihilation algebra for the transformed $y,q,x,p$ variables the commutation algebra given in (\ref{4.5g}) is replaced by one in which all commutators have positive signature \cite{Bender2008b}.

Also we note that while $\bar{H}^{\prime}$ is a straightforward free theory two-oscillator Hamiltonian, interactions are not as straightforward. If we add a $\lambda z^4$ term onto  $H_{\rm PU}(\omega_1,\omega_2)$ given in (\ref{4.3g}) this is equivalent to adding a term $\lambda y^4$ onto $\bar{H}$  given in (\ref{5.3g}). Thus in $\bar{H}+\lambda y^4$ the $\bar{H}$ term is not Hermitian while the $\lambda y^4$ term is. On the other hand if we now make the transformation to the Hermitian $\bar{H}^{\prime}$ given in (\ref{5.8e}) we have to transform the $\lambda y^4$ term as well.  On transforming $y$ we obtain \cite{Bender2008b}
\begin{eqnarray}
y^{\prime}=e^{-Q/2}ye^{Q/2}=y \cosh \theta+\frac{i}{\omega_1\omega_2}p\sinh\theta, \quad \tanh\theta=\frac{\omega_2}{\omega_1}.
\label{5.13e}
\end{eqnarray}
Thus now it is $\lambda y^{\prime 4}$ that is not Hermitian. Thus in neither case is the total Hamiltonian Hermitian. As we now show, this has consequences for interacting processes.

\section{Interactions}
\label{S6}
\begin{figure}[hptb]
\centering
 \includegraphics[scale=0.7]{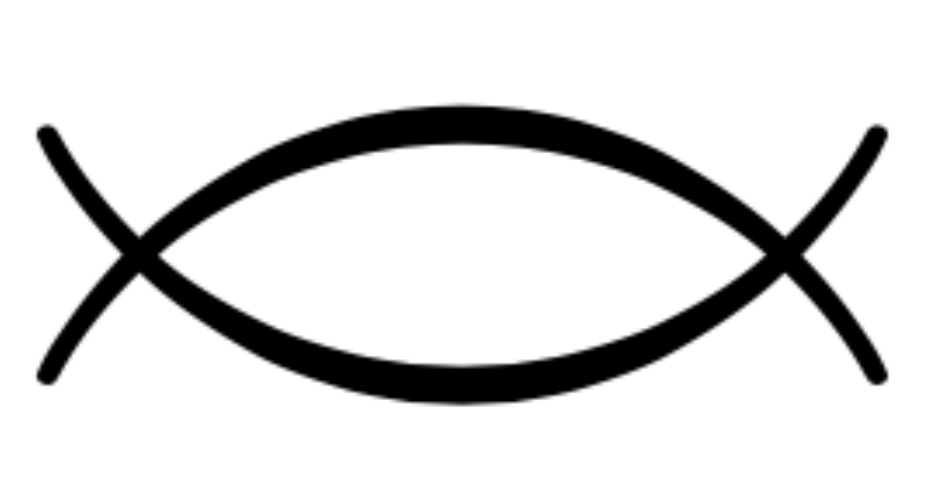}
\caption{First radiative correction in scalar field theory.} 
\label{between}
\end{figure}

Once we have shown that there are no negative norms associated with the free scalar-field theory action $I_S$ given in (\ref{3.6h}), it must be the case that when we add on an interaction such as $-\lambda \phi^4$ to $I_S$ no negative norm states can be generated by radiative corrections since one cannot change the signature of a Hilbert space in perturbation theory. Despite this, if we insert the propagator given in (\ref{3.8h}) into the one loop Feynman diagram shown in Fig. (\ref{between}) and  take the vertices to be of the $L({\rm interacting})=-\lambda \phi^4$ form, then because of the relative minus sign in (\ref{3.8h}) some of the discontinuities in the graph will be negative, and thus would on their own violate unitarity. Explicit calculation shows that the total discontinuity is given by \cite{Mannheim2018b}
 \begin{eqnarray}
 \Pi(p_0)=&-&\frac{i\lambda^2}{8\pi(M_1^2-M_2^2)^2}\bigg{(}\theta(p_0-2M_2)\frac{(p_0^2-4M_2^2)^{1/2}}{2p_0}+\theta(p_0-2M_1)\frac{(p_0^2-4M_1^2)^{1/2}}{2p_0}
 \nonumber\\
&-&\theta(p_0-M_1-M_2)\frac{[p_0^2-(M_1+M_2)^2]^{1/2}[p_0^2-(M_1-M_2)^2]^{1/2}}{p_0^2}\bigg{)},
\label{6.1g}
 \end{eqnarray}
where the total incoming four-vector is $(p_0,0,0,0)$. As we can see from (\ref{6.1g}), the contributions with positive discontinuity cannot cancel those with negative discontinuity since the various discontinuities start at different thresholds and have different mass dependences. 

However, this is not the complete situation, though it would have been had (the relativistic generalization of) either $\bar{H}+\lambda y^4$ or  $\bar{H}^{\prime}+\lambda y^{\prime 4}$ been Hermitian. But neither is. Nonetheless, since both have a $PT$ structure, they can be brought to a Hermitian form by the application of yet another similarity transformation of the form $S^{\prime}=\exp(-Q^{\prime}/2)$, so that the operator that  is now needed for the $V$-norm is  $V^{\prime}=\exp(-Q^{\prime})$. On doing this we change the conjugate of a right eigenvector $|R\rangle$ to $\langle L|=\langle R|\exp(-Q^{\prime})$, where the left eigenvector was introduced in Sec. \ref{S5}. In lowest order this $Q^{\prime}$ transformation will contain a term of order $\lambda$. Now the negative discontinuity given in (\ref{6.1g}) is of order $\lambda^2$. Thus to cancel it we need some other term of the same order. As noted in \cite{Mannheim2018b}, this term is provided by the tree approximation point vertex diagram (shaped like the letter $X$), since while it is nominally of order $\lambda$ it is evaluated between states that have been adjusted by  $V^{\prime}$. Since the modification is of order $\lambda$ the net contribution of the tree graph is of order $\lambda^2$. As shown in \cite{Mannheim2018b}, when this modified tree diagram is taken in conjunction with $\Pi(p_0)$ there is no net negative discontinuity in the scattering amplitude, and the net discontinuity is unitary.

To see why the $V$ modification is needed for unitarity, we note that the S-matrix is given not by $\sum |R_i\rangle\langle R_i|$ but by
\begin{eqnarray}
S=\sum_i |R_i\rangle\langle L_i|=\sum |R_i\rangle\langle R_i|V. 
\label{6.2g}
\end{eqnarray}
With the unitary time evolution operator being of the form $U=e^{-iHt}$, then with $H$ obeying $VHV^{-1}=H^{\dagger}$ as per (\ref{1.4g}), we find that 
\begin{eqnarray}
U^{\dagger}=e^{iH^{\dagger}t}=Ve^{iHt}V^{-1}=VU^{-1}V^{-1},
\label{6.3g}
\end{eqnarray}
so that instead of the standard (i.e., Hermitian-based) $UU^{\dagger}=I$ we have
\begin{eqnarray}
UV^{-1}U^{\dagger}V=I.
\label{6.4g}
\end{eqnarray}
As we see, the operator $V$ is central to unitarity considerations in scattering processes, and must always be taken into account in non-Hermitian theories that have an antilinear symmetry.

\section{Jordan-Block Realization of the Pais-Uhlenbeck Theory}
\label{S7}

Things change quite radically  when  we take the two frequencies to be equal. On setting  $\omega_1=\omega_2=\omega$, the $\omega$-dependent Hamiltonian given in (\ref{2.16g}) becomes the Hamiltonian given in  (\ref{2.15g}). Applying the frequency-independent (\ref{5.2g}) to (\ref{2.15g}) then brings it to the form
\begin{eqnarray}
e^{\pi p_zz/2}H_{\rm PU}(\omega)e^{-\pi p_zz/2}=\frac{p^2}{2}-iqx+\omega^2x^2+\frac{1}{2}\omega^4y^2.
\label{7.1g}
\end{eqnarray}
As previously the right-eigenvectors and left-eigenvectors are distinct, and for the ground state with energy $\omega$ the eigenfunctions given in (\ref{5.5e}) take the form \cite{Bender2008b}
\begin{eqnarray}
\psi^R_0(y,x,t)&=&\exp\left[-\omega^3y^2-\omega^2yx-\omega x^2-i\omega t\right],
\nonumber\\
\psi^L_0(y,x,t)&=&\exp\left[-\omega^3y^2+\omega^2yx-\omega x^2+i\omega t\right],
\label{7.2g}
\end{eqnarray}
with finite, positive, time-independent normalization
\begin{eqnarray}
\int dydx\psi^L_0(y,x,t)\psi^R_0(y,x,t)=\frac{\pi}{2\omega^2}.
\label{7.3g}
\end{eqnarray}
As we see, the $\omega_1\neq\omega_2$ ground state wave function given in (\ref{5.5e}), being non-degenerate, continues directly into the $\omega_1\neq\omega_2$ ground state given in (\ref{7.3g}). 

However, this is not the case for the two one-particle states. Before we set $\omega_1=\omega_2$ the two one-particle excited states with  eigenvalues $3\omega_1/2+\omega_2/2$ and $\omega_1/2+3\omega_2/2$ have wave functions and normalizations that take the form \cite{Bender2008b} 
\begin{eqnarray}
\psi^R_1(y,x,t)&=&(x+\omega_2y){\rm exp}\left[-\tfrac{1}{2}(\omega_1+\omega_2)\omega_1\omega_2
y^2-\omega_1\omega_2yx-\tfrac{1}{2}(\omega_1+\omega_2)x^2-\tfrac{i}{2}(3\omega_1+\omega_2)t\right],
\nonumber\\
\psi^L_1(y,x,t)&=&\eta_1(x-\omega_2y){\rm exp}\left[-\tfrac{1}{2}(\omega_1+\omega_2)\omega_1\omega_2
y^2+\omega_1\omega_2yx-\tfrac{1}{2}(\omega_1+\omega_2)x^2+\tfrac{i}{2}(3\omega_1+\omega_2)t\right],
\nonumber\\
\psi^R_2(y,x,t)&=&(x+\omega_1y){\rm exp}\left[-\tfrac{1}{2}(\omega_1+\omega_2)\omega_1\omega_2
y^2-\omega_1\omega_2yx-\tfrac{1}{2}(\omega_1+\omega_2)x^2-\tfrac{i}{2}(\omega_1+3\omega_2)t\right],
\nonumber\\
\psi^L_2(y,x,t)&=&\eta_2(x-\omega_1y){\rm exp}\left[-\tfrac{1}{2}(\omega_1+\omega_2)\omega_1\omega_2
y^2+\omega_1\omega_2yx-\tfrac{1}{2}(\omega_1+\omega_2)x^2+\tfrac{i}{2}(\omega_1+3\omega_2)t\right],
\nonumber\\
&&\int dydx\psi^L_1(y,x,t)\psi^R_1(y,x,t)=\frac{\eta_1\pi(\omega_1-\omega_2)}{2(\omega_1+\omega_2)^2\omega_1^{3/2}\omega_2^{1/2}},
\nonumber\\
&&\int dydx\psi^L_2(y,x,t)\psi^R_2(y,x,t)=-\frac{\eta_2\pi(\omega_1-\omega_2)}{2(\omega_1+\omega_2)^2\omega_1^{1/2}\omega_2^{3/2}},
\label{7.4g}
\end{eqnarray}
where the phases $\eta_1=1$, $\eta_2=-1$ are fixed in accordance with (\ref{5.10e}).
In the limit both energies become  $2\omega$, and the wave functions collapse onto the common \cite{Bender2008b} 
\begin{eqnarray}
\psi^R_1(y,x,t)&=&(x+\omega y)\exp\left[-\omega^3y^2-\omega^2yx-\omega x^2-2i\omega t\right],
\nonumber\\
\psi^L_1(y,x,t)&=&\eta_1(x-\omega y)\exp\left[-\omega^3y^2+\omega^2yx-\omega x^2+2i\omega t\right].
\label{7.5g}
\end{eqnarray}
Thus only one wave function survives, and the Hamiltonian becomes of non-diagonalizable Jordan-block form. As is characteristic of Jordan-block Hamiltonians the norm becomes zero with 
\begin{eqnarray}
\int dydx\psi^L_1(y,x,t)\psi^R_1(y,x,t)=0.
\label{7.6g}
\end{eqnarray}

To see what happened to the other wave functions we set $\omega_1=\omega +\epsilon$, $\omega_2=\omega -\epsilon$, and on taking the $\epsilon \rightarrow 0$ limit obtain 
\begin{eqnarray}
\psi^R_{1a}(y,x,t)&=&\lim_{\epsilon \rightarrow 0}\frac{1}{2\epsilon}(\psi^R_2(y,x,t)-\psi^R_1(y,x,t))=[(x+\omega y)it+y]\exp\left[-\omega^3y^2-\omega^2yx-\omega x^2-2i\omega t\right],
\nonumber\\
\psi^L_{1a}(y,x,t)&=&\lim_{\epsilon \rightarrow 0}\frac{1}{2\epsilon}(\eta_2\psi^L_2(y,x,t)-\eta_1\psi^L_1(y,x,t))=-[(x-\omega y)it+y]\exp\left[-\omega^3y^2+\omega^2yx-\omega x^2+2i\omega t\right],
\label{7.7g}
\end{eqnarray}
with the resulting linear in $t$ dependence indicating that these two wave functions are not stationary. 
Thus between (\ref{7.5g}) and (\ref{7.7g}) we still have the same number of wave functions that we had  in (\ref{7.4g}). These four wave functions still form a complete basis for the one-particle sector of the equal frequency Hilbert space and its dual space. It is just the eigenbasis of the equal frequency Hamiltonian that is not complete.

Even though $\psi^R_{1a}(y,x,t)$ and $\psi^L_{1a}(y,x,t)$ are not eigenstates of the equal-frequency Hamiltonian, they are solutions to the time-dependent Schr\"odinger equation. Then, since they have the same good asymptotic behavior as the eigenstates themselves one can still construct time-independent probabilities for them.  These are of the form \cite{Bender2008b}
\begin{eqnarray}
\int dydx\psi^L_{1}(y,x,t)\psi^R_{1a}(y,x,t)=\int dydx\psi^L_{1a}(y,x,t)\psi^R_{1}(y,x,t)=- \frac{\pi}{8\omega^4},
\nonumber\\
\int dydx\psi^L_{1a}(y,x,t)\psi^R_{1a}(y,x,t)=- \frac{\pi}{8\omega^5}.
\label{7.8g}
\end{eqnarray}
Now initially it does not appear to be possible that these probabilities could be time independent since both $\psi^R_{1a}(y,x,t)$ and $\psi^L_{1a}(y,x,t)$ are growing in time. And yet they have to be since they both obey the Schr\"odinger equation, and the time derivatives of these probabilities can be written as asymptotic spatial surface terms that vanish. Mathematically, the reason why these  probabilities are time independent is because each term that grows with $t$ is multiplied by a coefficient that is zero since the states in (\ref{7.6g})  have zero norm. With this same pattern repeating for the higher excited states as well, and with the stationary and non-stationary states together forming a complete basis, any wave packet that is built out of them will preserve its probability in time. Despite it being radically different from the way unitarity is implemented for Hamiltonians that have a complete basis of eigenstates, this is how unitarity is achieved for non-diagonalizable, Jordan-block Hamiltonians. And as such, it is completely viable quantum-mechanically, showing that quantum mechanics is far richer than conventional Hermitian quantum mechanics  with Dirac inner products. 

To gain more insight into this analysis we consider the wave equations given in (\ref{3.7h}) and (\ref{4.2h}). As long as $\omega_1\neq \omega_2$  the solutions to  (\ref{4.2h}) are of the form $\exp(-i\omega_1t)$, $\exp(-i\omega_2t)$ and their conjugates. However, when we set $\omega_1=\omega_2=\omega$ (or analogously $M_1=M_2=M$), the solutions take a different form, viz.   $\exp(-i\omega_1t)$, $t\exp(-i\omega_1t))$ and their conjugates, since
\begin{eqnarray}
(\partial_t^2+\omega^2) (\partial_t^2+\omega^2)(ae^{-i\omega t}+bte^{-i\omega t})=(\partial_t^2+\omega^2) [-2ib\omega e^{-i\omega t}]=0.
\label{7.9g}
\end{eqnarray}
Thus the second class of solutions possesses a term that is linear in $t$. Since such a term is not stationary, already in the classical wave equation prior to quantization we can see that the quantized version of the theory will be Jordan block.  However, we should also note that in the classical theory, while the appropriate fourth-order theory Hamiltonian may have an Ostrogradski instability, it is still a time independent constant of motion in these solutions (since that is what Ostrogradski set out to construct), so there is no runaway in energy. And even in the classical theory, if formulated with pure imaginary $z$ there is no instability. That we can continue coordinates into the complex plane even in the classical theory is allowed since symplectic transformations with complex angles preserve the Poisson bracket algebra. In fact, the complex similarity transformations that we made in the quantum theory parallel complex symplectic transformations in the classical theory. And if they both transform through the same angles, there is even a correspondence principle at each of those angles, with Poisson brackets mapping into commutators \cite{Mannheim2018a}.

We conclude this section by noting three additional features of the equal-frequency limit. First we note that in the limit both $D_{\rm PU}(k)$ given in (\ref{3.8h}) and $G_{\rm PU}(E)$ given in (\ref{4.2h}) become singular. Consequently, the  partial fraction decomposition breaks down, and any insight that it might provide into the structure of the eigenspectrum does not carry over to the equal-frequency case. Secondly, the canonical commutator structure given in (\ref{4.5g}) also becomes singular in the limit. And thirdly, the similarity transformation required to diagonalize $H_{\rm PU}$ as given in (\ref{5.7e}) and (\ref{5.8e})  becomes singular in the limit as well, so that no diagonalization could be effected. The equal-frequency theory thus represents a whole branch of quantum mechanics that is simply inaccessible with Hermitian Hamiltonians. As we will see in Sec. \ref{S9} this could play a central role in constructing a viable quantum theory of gravity.

\section{Complex Conjugate Energy Pair Realization of Pais-Uhlenbeck Theory}
\label{S8}

For the complex conjugate energy pair realization with $\omega_1=\alpha+i\beta$,  $\omega_2=\alpha-i\beta$ with real and positive $\alpha$ and $\beta$ we apply the similarity transform given in (\ref{5.2g}) to (\ref{2.14g}) and obtain
\begin{eqnarray}
e^{\pi p_zz/2}H_{\rm PU}(\alpha,\beta)e^{-\pi p_zz/2}=\frac{p}{2}-iqx+(\alpha^2-\beta^2)x^2+\frac{1}{2}(\alpha^2+\beta^2)^2y^2.
\label{8.1g}
\end{eqnarray}
The ground state energy $(\omega_1+\omega_2)/2=\alpha$ is still real, and the associated eigenfunctions are of the form
\begin{eqnarray}
\psi^R_0(y,x,t)&=&{\rm exp}\left[-\alpha(\alpha^2+\beta^2)
y^2-(\alpha^2+\beta^2)yx-\alpha x^2-i\alpha t\right],
\nonumber\\
\psi^L_0(y,x,t)&=&{\rm exp}\left[-\alpha(\alpha^2+\beta^2)
y^2+(\alpha^2+\beta^2)yx-\alpha x^2+i\alpha t\right],
\label{8.2g}
\end{eqnarray}
with a normalization
\begin{eqnarray}
\int dx dy\psi^L_0(y,x,t)\psi^R_0(y,x,t)=\frac{\pi}{2\alpha(\alpha^2+\beta^2)^{1/2}}
\label{8.3g}
\end{eqnarray}
that is finite, time-independent and real. 

The one-particle state wave functions are
\begin{eqnarray}
\psi^R_1(y,x,t)&=&(x+(\alpha-i\beta)y){\rm exp}\left[-\alpha(\alpha^2+\beta^2)
y^2-(\alpha^2+\beta^2)yx-\alpha x^2-i(2\alpha+i\beta)t\right],
\nonumber\\
\psi^L_1(y,x,t)&=&(x-(\alpha-i\beta)y){\rm exp}\left[-\alpha(\alpha^2+\beta^2)
y^2+(\alpha^2+\beta^2)yx-\alpha x^2+i(2\alpha+i\beta)t\right],
\nonumber\\
\psi^R_2(y,x,t)&=&(x+(\alpha+i\beta)y){\rm exp}\left[-\alpha(\alpha^2+\beta^2)
y^2-(\alpha^2+\beta^2)yx-\alpha x^2-i(2\alpha-i\beta)t\right],
\nonumber\\
\psi^L_2(y,x,t)&=&(x-(\alpha+i\beta)y){\rm exp}\left[-\alpha(\alpha^2+\beta^2)
y^2+(\alpha^2+\beta^2)yx-\alpha x^2+i(2\alpha-i\beta)t\right].
\label{8.4g}
\end{eqnarray}
As can be shown, and as follows from (\ref{1.5g}), the overlaps $\int dydx\psi^L_1(y,x,t)\psi^R_1(y,x,t)=i\pi \beta (\alpha-i\beta)/(4\alpha^2(\alpha^2+\beta^2)^{3/2})$ and $\int dydx\psi^L_2(y,x,t)\psi^R_2(y,x,t)=-i\pi \beta (\alpha+i\beta)/(4\alpha^2(\alpha^2+\beta^2)^{3/2})$ are both finite and time independent. And the overlaps $\int dydx\psi^L_2(y,x,t)\psi^R_1(y,x,t)$ and $\int dydx\psi^L_1(y,x,t)\psi^R_2(y,x,t)$ where the time dependences would not cancel are both zero on account of the spatial integrations.

It should be emphasized that the $\int dydx\psi^L_1(y,x,t)\psi^R_1(y,x,t)$ and $\int dydx\psi^L_2(y,x,t)\psi^R_2(y,x,t)$ overlaps are time independent. Thus despite the existence of states that both grow and decay exponentially in time one can still construct transition matrix elements that are constant in time. And in fact this is enforced by (\ref{1.5g}), since in the complex energy pair realization of the theory (\ref{1.4g}) still holds. The pseudo-Hermitian condition $VHV^{-1}=H^{\dagger}$ thus takes care of the time dependence of matrix elements.  This should be contrasted with the standard discussion of decaying states. There one only allows for exponentials that decay, and one does so by actually taking the Hamiltonian to have a non-Hermitian component. One does not allow for exponentials that grow. In the antilinear case exponential behavior is natural, and not only are both growing and decaying exponentials present, they regulate each other to give time independent transitions. That these transitions are time independent is due to the fact that the decaying system and the system that it decays into are both involved, with a decrease in population of the decaying system necessitating an increase in population of the system into which it decays. The treatment of decays in the antilinear case is thus more comprehensive than the standard treatment.

One additional interesting feature of $H_{\rm PU}(\alpha,\beta)$ is that it contains no ad hoc dissipative term of the type one uses for the decay of a single harmonic oscillator, and yet still describes decays. This is because not only is the decaying system taken into consideration, the decay products are included in the analysis too, i.e., a two-oscillator system.

\section{Implications for Gravity}
\label{S9}

As a quantum theory the standard second-order derivative Einstein gravitational theory is not renormalizable since its $1/k^2$ propagator generates uncontrollable infinities in the ultraviolet. To attempt to ameliorate this difficulty one can add on terms that are quadratic in the Ricci tensor $R_{\mu\nu}$ and the Ricci scalar $R^{\alpha}_{\phantom{\alpha}\alpha}$, to give an action of the form
\begin{eqnarray}
I_{\rm GRAV}=\int d^4x(-g)^{1/2}\left[-AR^{\alpha}_{\phantom{\alpha}\alpha}+B(R^{\alpha}_{\phantom{\alpha}\alpha})^2+CR^{\mu\nu}R_{\mu\nu}\right],
\label{9.1g}
\end{eqnarray}
where $A$, $B$ and $C$ are constants, with $A$ having the dimension of inverse length squared, and $B$ and $C$ being dimensionless. One has no need to include any $R_{\lambda\mu\nu\kappa}R^{\lambda\mu\nu\kappa}$ term as it can be traded for the $(R^{\alpha}_{\phantom{\alpha}\alpha})^2$ and $R^{\mu\nu}R_{\mu\nu}$ terms because $(-g)^{1/2}\left[R_{\lambda\mu\nu\kappa}
R^{\lambda\mu\nu\kappa}-4R_{\mu\kappa}R^{\mu\kappa}+(R^{\alpha}_{\phantom{\alpha}\alpha})^2\right]$ is a total divergence. 
On adding on a matter source with energy-momentum tensor $T_{\mu\nu}$, variation  of this action with respect to the metric generates a gravitational equation of motion of the form
\begin{eqnarray}
A G_{\mu\nu}+BW^{(1)}_{\mu\nu}+CW^{(2)}_{\mu\nu}=-\frac{1}{2}T_{\mu\nu}.
\label{9.2h}
\end{eqnarray}
Here $G_{\mu\nu}$ is the Einstein tensor and $W^{(1)}_{\mu\nu}$ and $W^{(2)}_{\mu\nu}$ may for instance be found in \cite{Mannheim2006}, with these various terms being of the form 
\begin{eqnarray}
G^{\mu\nu}&=&R^{\mu\nu}-\frac{1}{2}g^{\mu\nu}g^{\alpha\beta}R_{\alpha\beta},
\nonumber\\
W^{\mu \nu}_{(1)}&=&
2g^{\mu\nu}\nabla_{\beta}\nabla^{\beta}R^{\alpha}_{\phantom{\alpha}\alpha}                                             
-2\nabla^{\nu}\nabla^{\mu}R^{\alpha}_{\phantom{\alpha}\alpha}                          
-2 R^{\alpha}_{\phantom{\alpha}\alpha}R^{\mu\nu}                              
+\frac{1}{2}g^{\mu\nu}(R^{\alpha}_{\phantom{\alpha}\alpha})^2,
\nonumber\\
W^{\mu \nu}_{(2)}&=&
\frac{1}{2}g^{\mu\nu}\nabla_{\beta}\nabla^{\beta}R^{\alpha}_{\phantom{\alpha}\alpha}
+\nabla_{\beta}\nabla^{\beta}R^{\mu\nu}                    
 -\nabla_{\beta}\nabla^{\nu}R^{\mu\beta}                       
-\nabla_{\beta}\nabla^{\mu}R^{\nu \beta}                          
 - 2R^{\mu\beta}R^{\nu}_{\phantom{\nu}\beta}                                    
+\frac{1}{2}g^{\mu\nu}R_{\alpha\beta}R^{\alpha\beta}.
\label{9.3h}
\end{eqnarray}                                 
If we now linearize about  flat spacetime with background metric $\eta_{\mu\nu}$ and fluctuation  metric $g_{\mu\nu}=\eta_{\mu\nu}+h_{\mu\nu}$, to first perturbative order we obtain 
\begin{eqnarray}
\delta G^{\mu\nu}&=&\frac{1}{2}\left(\partial_{\alpha}\partial^{\alpha}h_{\mu\nu}-\partial_{\mu}\partial^{\alpha}h_{\alpha\nu}-\partial_{\nu}\partial^{\alpha}h_{\alpha\mu}+\partial_{\mu}\partial_{\nu}h\right)-\frac{1}{2}\eta_{\mu\nu}\left(\partial_{\alpha}\partial^{\alpha}h-\partial^{\alpha}\partial^{\beta}h_{\alpha\beta}\right),
\nonumber\\
\delta W^{(1)}_{\mu\nu}&=&[2\eta_{\mu\nu}\partial_{\alpha}\partial^{\alpha} -2\partial_{\mu}\partial_{\nu}]
[\partial_{\beta}\partial^{\beta}h-\partial_{\lambda}\partial_{\kappa}h^{\lambda\kappa}],
\nonumber\\
\delta W^{(2)}_{\mu\nu}&=&\frac{1}{2}\partial_{\beta}\partial^{\beta}[\partial_{\mu}\partial_{\nu}h
-\partial_{\mu}\partial_{\lambda}h^{\lambda}_{\phantom{\lambda}\nu}
-\partial_{\nu}\partial_{\lambda}h^{\lambda}_{\phantom{\lambda}\mu}
+\partial_{\lambda}\partial^{\lambda}h_{\mu\nu}]
+\frac{1}{2}[g_{\mu\nu}\partial_{\beta}\partial^{\beta} -2\partial_{\mu}\partial_{\nu}][\partial_{\lambda}\partial^{\lambda}h-\partial_{\kappa}\partial_{\lambda}h^{\kappa\lambda}].
\label{9.4h}
\end{eqnarray}                                 
Here $h=\eta^{\mu\nu}h_{\mu\nu}$, and $\delta W^{\mu \nu}_{(1)}$ and $\delta W^{\mu \nu}_{(2)}$ are given in \cite{Mannheim2012b}. On taking the trace of the fluctuation around a background (\ref{9.2h}) we obtain 
\begin{eqnarray}
[-3A+(6B+2C)\partial_{\beta}\partial^{\beta}]\left(\partial_{\lambda}\partial^{\lambda}h-\partial_{\kappa}\partial_{\lambda}h^{\kappa\lambda}\right)=-\frac{1}{2}\eta^{\mu\nu}\delta T_{\mu\nu}.
\label{9.5h}
\end{eqnarray}
In the convenient transverse gauge where $\partial_{\mu}h^{\mu\nu}=0$, the propagator for $h$ is given by
\begin{eqnarray}
D(h,k^2)=\frac{1}{(6B+2C)k^2(k^2-M^2)}=-\frac{1}{M^2(6B+2C)}\left(\frac{1}{k^2}-\frac{1}{k^2-M^2}\right),\quad M^2=\frac{-3A}{6B+2C}.
\label{9.6h}
\end{eqnarray}

As we see, in this case the graviton propagator for $h$ that would be associated with the Einstein tensor $\delta G_{\mu\nu}$ alone  is replaced by the Pauli-Villars type $D_{\rm PV}(k)=1/k^2-1/(k^2-M^2)$ propagator given in (\ref{3.1g}). And now the leading behavior at large momenta is $1/k^4$. In consequence, the theory is now renormalizable \cite{Stelle1977}. However, as such the theory suffers some shortcomings, the possibility of ghosts due to the relative minus sign in $D_{\rm PV}(k)$ or the possibility of negative energies, and the existence of additional particles due to the presence of poles in the $1/(k^2-M^2)$ component of the propagator \cite{Stelle1977}. That there would be additional particles in the theory is actually an indirect consequence of a theorem due to Weinberg \cite{Weinberg1965}, namely that if one has a single massless spin two particle in a standard positive definite Hilbert space it must couple via the Einstein tensor. Thus any departure from pure Einstein would necessitate new particles (string theory also has this concern).  As we had noted above, one can avoid negative energies as they lie in a different Hilbert space than the ghost states. And then one can even avoid negative Dirac norm ghost states since one instead has to use  the $PT$ symmetry theory inner product. The work of \cite{Bender2008a} thus makes the theory based on (\ref{9.1g}) unitary. However, one cannot get rid of the extra states, and making them heavy enough to have evaded detection may only be postponing the inevitable. 

There is however an alternate fourth-order derivative theory that is renormalizable, has no negative energy particles, is unitary, and yet does not have additional particles, namely the conformal gravity theory that has been advocated and explored in \cite {Mannheim2006,Mannheim2012,Mannheim2017} and references therein. In the theory the requirement of local conformal invariance under $g_{\mu\nu}(x)\rightarrow e^{2\alpha(x)}g_{\mu\nu}(x)$, where $\alpha(x)$ is a local function of the coordinates, requires a polynomial action to have the unique form 
\begin{eqnarray}
I_{\rm W}=-\alpha_g\int d^4x\, (-g)^{1/2}C_{\lambda\mu\nu\kappa}
C^{\lambda\mu\nu\kappa}
\equiv -2\alpha_g\int d^4x\, (-g)^{1/2}\left[R_{\mu\kappa}R^{\mu\kappa}-\frac{1}{3} (R^{\alpha}_{\phantom{\alpha}\alpha})^2\right].
\label{9.7h}
\end{eqnarray}
Here $\alpha_g$ is a dimensionless  gravitational coupling constant, and $C_{\lambda\mu\nu\kappa}$ is the conformal Weyl tensor. Functional variation with respect to the metric $g_{\mu\nu}(x)$ generates fourth-order derivative gravitational equations of motion of the form (see e.g. \cite{Mannheim2006})
\begin{eqnarray}
\frac{1}{(-g)^{1/2}}\frac{\delta I_{\rm W}}{\delta g_{\mu\nu}}=-2\alpha_g W^{\mu\nu}=-2\alpha_g\left[2\nabla_{\kappa}\nabla_{\lambda}C^{\mu\lambda\nu\kappa}-
R_{\kappa\lambda}C^{\mu\lambda\nu\kappa}\right]=-2\alpha_g\left[W^{\mu
\nu}_{(2)}-\frac{1}{3}W^{\mu\nu}_{(1)}\right]=-\frac{1}{2}T^{\mu\nu},
\label{9.8h}
\end{eqnarray}
where the functions $W^{\mu \nu}_{(1)}$ and $W^{\mu \nu}_{(2)}$ are given above.

As such, the theory admits of a vacuum solution in which the Ricci tensor vanishes, and is thus able to meet the same solar system tests as the Ricci flat solution to Einstein gravity. However, the conformal theory has other solutions beyond the Newtonian potential, such as linear and quadratic potentials, and is then able to fit galactic rotation curves and the accelerating universe data without dark matter and with a cosmological constant that is under control because of the underlying conformal symmetry of the theory (see \cite{Mannheim2017} and references therein). 

For our purposes here we note that when $W_{\mu\nu}$ is linearized about a flat background, to first perturbative order one obtains \cite{Mannheim2011}
\begin{eqnarray}
\delta W_{\mu\nu}=\frac{1}{2}(\eta^{\rho}_{\phantom{\rho} \mu} \partial^{\alpha}\partial_{\alpha}-\partial^{\rho}\partial_{\mu})
(\eta^{\sigma}_{\phantom{\sigma} \nu} \partial^{\beta}\partial_{\beta}-
\partial^{\sigma}\partial_{\nu})K_{\rho \sigma}- 
\frac{1}{6}(\eta_{\mu \nu} \partial^{\gamma}\partial_{\gamma}-
\partial_{\mu}\partial_{\nu})(\eta^{\rho \sigma} \partial^{\delta}\partial_{\delta}-
\partial^{\rho}\partial^{\sigma})K_{\rho\sigma},
\label{9.9h}
\end{eqnarray}
where $K_{\mu\nu} =h_{\mu\nu}-\tfrac{1}{4}\eta_{\mu\nu}\eta^{\alpha\beta}h_{\alpha\beta}$ with $\eta^{\mu\nu}K_{\mu\nu}=0$. Now we had shown above that both $\delta W^{\mu \nu}_{(1)}$ and $\delta W^{\mu \nu}_{(2)}$ depend on the trace $h$. However, in the particular linear combination $\delta W^{\mu \nu}_{(2)}-(1/3) \delta W^{\mu \nu}_{(1)}$ the trace drops out and $\delta W^{\mu \nu}$ only depends on the traceless $K_{\mu\nu}$. The distinction here is that  in (\ref{9.1g}) the $B(R^{\alpha}_{\phantom{\alpha}\alpha})^2$ and $CR^{\mu\nu}R_{\mu\nu}$ terms with dimensionless $B$ and $C$ are separately globally scale invariant, i.e., invariant under $g_{\mu\nu}(x)\rightarrow e^{2\alpha}g_{\mu\nu}(x)$ where $\alpha$ is a constant. However, the action based on the particular combination $R^{\mu\nu}R_{\mu\nu}-(1/3)(R^{\alpha}_{\phantom{\alpha}\alpha})^2$ has a local symmetry for all local $\alpha(x)$.
With $W_{\mu\nu}$ being traceless, as is appropriate to a conformal theory, it follows that $\delta [\eta^{\mu\nu}W_{\mu\nu}]=\eta^{\mu\nu}\delta W_{\mu\nu}-h^{\mu\nu}W_{\mu\nu}=0$. But for a flat background $W_{\mu\nu}=0$, and thus for fluctuations around flat it follows that, as can readily be checked,  $\eta^{\mu\nu}\delta W_{\mu\nu}=0$. 
Thus even though there are ten $h_{\mu\nu}$ fluctuations, because $\delta W_{\mu\nu}$  is traceless it can only depend on the nine degrees contained in the traceless $K_{\mu\nu}$. 

From the structure of  (\ref{9.9h}), we find that in the convenient  transverse gauge $\partial_{\mu}K^{\mu\nu}=0$ the perturbative equation of motion (\ref{9.9h}) simplifies to  \cite{Mannheim2011}
\begin{eqnarray}
4\alpha_g \delta W_{\mu\nu}=2\alpha_g(\partial_t^2-\vec{\nabla}^2)^2K_{\mu\nu}=\delta T_{\mu\nu}.
\label{9.10h}
\end{eqnarray}
With the perturbative equation of motion being diagonal in the $(\mu,\nu)$ indices, we can treat each component of $K_{\mu\nu}$ independently.
Noting the similarity with (\ref{7.9g}), we see that the solutions to a source-free (\ref{9.10h}) are of the form $\exp(-ik\cdot x)$ and  $t\exp(-ik\cdot x)$ together with their conjugates. As our discussion in Sec. \ref{S7} shows, with there being such non-stationary solutions, the quantum Hamiltonian of the theory must be of Jordan-block form, and in \cite{Mannheim2011} this was confirmed explicitly. In momentum space the propagator associated with (\ref{9.10h}) behaves as $1/k^4$, to thus have power counting renormalizable behavior in the ultraviolet. This $1/k^4$ propagator is often thought of as being a limit of a massive particle Pauli-Villars propagator, but as we have shown above this cannot be the case since the limit of equal masses or of setting $A=0$, i.e., $M^2=0$, in (\ref{9.6h}) is singular. Moreover, we note that the $D(h,k^2)$ propagator given in (\ref{9.6h}) becomes undefined if $6B++2C=0$. This is precisely the conformal gravity case where $B=-C/3$, and this has to be so since the trace $h$ decouples from the conformal gravity fluctuation equation given in (\ref{9.9h}). Thus as we see, not only is the limit from second order plus fourth order to pure fourth order  singular, the limit from a globally scale invariant fourth-order theory to a locally conformal invariant theory is singular too.

To the extent that one might want to relate the $1/k^4$  propagator to a limit of a propagator with mass we could write it as
\begin{eqnarray}
\frac{1}{(k^2+i\epsilon)^2}=\lim_{M^2\rightarrow 0} \frac{d}{dM^2}
\left(\frac{1}{k^2-M^2+i\epsilon}\right).
\label{9.11h}
\end{eqnarray}
This shows that the number of poles  is the same as the number of poles that occur in a second-order theory rather than in a massive fourth-order theory, while also showing that there are no relative minus signs in the propagator. The particle spectrum of pure fourth-order theories is totally different from that of a second-order plus fourth-order theory. Thus even while conformal gravity is based on fourth-order derivative equations it does not have a proliferation of particle states. First, it loses particle states since some of the states are not eigenstates, and second the eigenstates that it does have are zero norm states that would leave no imprint in a detector. Even though the theory only contains one graviton eigenstate despite not being based on the Einstein equations, it evades Weinberg's theorem by not being formulated in a positive definite Hilbert space but in one with zero norms. Since (\ref{9.10h}) is a wave equation, conformal gravity does have gravitational waves even as the quantized form of these waves has zero norm. Quantum conformal gravity  is renormalizable, is unitary in the way described in Sec. \ref{S7} for Jordan-block systems, and has no extraneous particles. All of this is achievable because conformal gravity is a $PT$ theory rather than an Hermitian one, and could not have  been achieved otherwise. 

Beyond this there is one further aspect to the theory, which initially might appear to be of concern. Since we need to make the transformation in (\ref{5.2g}) we would be led to a metric of the form $ig_{\mu\nu}$. However, since $g^{\mu\nu}g_{\nu\tau}=\delta^{\mu}_{\tau}$, we would have to make a change $-ig^{\mu\nu}$ in the contravaraint component of the metric. Then, since the Levi-Civita connection $\Gamma^{\lambda}_{\mu\nu}=\tfrac{1}{2}g^{\lambda\alpha}(\partial_{\mu}g_{\nu\alpha} +\partial_{\nu}g_{\mu\alpha}-\partial_{\alpha}g_{\nu\mu} )$ contains both contravariant and covariant metric components, it would be left unchanged under this transformation. Since the Riemann tensor $R^{\lambda}_{\phantom{\lambda}\mu\nu\kappa}$ is built out of the connection and its derivatives the Riemann tensor is also not affected by the change in phase of the metric, though it could be affected by raising or lowering indices. Then, with (\ref{9.7h})  being quadratic in the Riemann tensor the $I_{\rm W}$ action would still be real.

Now a reader might think that higher-derivative quantum gravity is only of academic interest.  To see that this is not the case consider the Dirac action for a massless fermion coupled to a background geometry of the form
\begin{eqnarray}
I_{\rm D}=\int d^4x(-g)^{1/2}i\bar{\psi}\gamma^{c}V^{\mu}_c(\partial_{\mu}+\Gamma_{\mu})\psi, 
\label{9.12h}
\end{eqnarray}
where the $V^{\mu}_a$ are vierbeins and $\Gamma_{\mu}=-(1/8)[\gamma_a,\gamma_b](V^b_{\nu}\partial_{\mu}V^{a\nu}+V^b_{\lambda}\Gamma^{\lambda}_{\nu\mu}V^{a\nu})$ is the spin connection as introduced in order to make $I_{\rm D}$ be locally Lorentz invariant. 
Now while the spin connection  $\Gamma_{\mu}$ was not at all introduced for the purpose, in its presence $I_{\rm D}$ is  locally conformal invariant under 
\begin{eqnarray}
V^{\mu}_a\rightarrow e^{-\alpha(x)}V^{\mu}_a(x),\quad \psi(x)\rightarrow e^{-3\alpha(x)/2}\psi(x),\quad g_{\mu\nu}(x)\rightarrow e^{2\alpha(x)} g_{\mu\nu}(x).
\label{9.13h}
\end{eqnarray} 
We thus get local conformal invariance for free. In fact, other than the double-well Higgs potential, the entire $SU(3)\times SU(2)\times U(1)$ standard model is locally conformal invariant. Thus if fermion masses are generated dynamically \cite{Mannheim2017}, then the entire standard model would be locally conformal invariant. 't Hooft \cite{tHooft2015b} has also argued that there should be an underlying local conformal symmetry in nature.

We now introduce the path integral $\int D[\psi]D[\bar{\psi}]\exp{iI_{\rm D}}=\exp(iI_{\rm EFF})$, and on performing the path integration on $\psi$ and $\bar{\psi}$ obtain an effective action with leading term  \cite{tHooft2010a}
\begin{eqnarray}
I_{\rm EFF}&=&\int d^4x(-g)^{1/2}\frac{C}{20}\left[R_{\mu\nu}R^{\mu\nu}-\tfrac{1}{3}(R^{\alpha}_{\phantom{\alpha}\alpha})^2\right],
\label{9.14h}
\end{eqnarray}
where $C$ is a log divergent constant. Thus when the standard model is coupled to gravity we generate the fourth-order derivative conformal gravity action. But the standard model is ghost free. Since this particular fermion path integration is equivalent to a one loop Feynman diagram, and since one cannot change the signature of a Hilbert space in perturbation theory, conformal gravity must be ghost free too. And if it were not, then the standard model would not be unitary either. When coupled to gravity, radiative corrections to $SU(3)\times SU(2)\times U(1)$  generate fourth-order conformal gravity whether we like it or not. We thus have to deal with conformal gravity one way or another, and we do deal with it using the $PT$ symmetry inner product. Now the steps leading from the flat space Dirac action to the curved space (\ref{9.12h}) and then to (\ref{9.14h}) are all completely standard. Thus the steps leading to the conformal gravity action given in (\ref{9.14h}) are all completely standard and beyond reproach. But conformal gravity is a $PT$ theory. We are thus led, essentially inexorably,  to gravity being $PT$ theory, and to $PT$ theory being relevant to fundamental physics.

\section{Final Comments}
\label{S10}

In this paper we have explored Hamiltonians that are not Hermitian but instead have an antilinear symmetry. For these theories to give time independent probabilities we have had to replace the standard Dirac inner product by the antilinear symmetry inner product in which the bra vector is no longer the Hermitian conjugate of the ket vector but is instead its antilinear symmetry conjugate. That we can do this at all is because the Schr\"odinger equation only involves the ket vectors. There is thus some freedom in picking the appropriate bra vectors for the dual space. This freedom has been present in quantum mechanics since its inception but had been overlooked for many years until the emergence of the $PT$ symmetry program. This modification of the inner product is not in any way a change in quantum mechanics, it only uses what quantum mechanics had always allowed, and shows how rich it can actually be. Antilinear symmetry has three distinct realizations: eigenvalues all real and eigenvectors complete, eigenvalues all real but eigenvectors incomplete, eigenvalues in complex conjugate pairs and eigenvectors complete. Of these, only the first one could be realized by Hermitian operators, with the latter two providing options for quantum mechanics that go beyond the Hermitian case. While it had not at all been developed for the purpose, the change in the inner product allows for a solution to the negative ghostlike Dirac inner product that had plagued higher-derivative quantum field theories. Should it turn out that quantum gravity is associated with a fourth-order derivative theory, that theory would have to be a $PT$ theory rather than a Hermitian one. If one of the four fundamental forces in nature turns out to be a $PT$ theory that would constitute a quite considerable achievement for the $PT$ symmetry program.

\end{document}